\documentclass[12pt]{iopart}
\usepackage{iopams}  
\usepackage{bbm}  
\usepackage{epsfig}  

\usepackage{color}
\definecolor{blue}{rgb}{0,0,1}

\newcommand{\text}[1]{{\rm #1}}
\newcommand{\eqref}[1]{(\ref{#1})}

\begin{document}

\title{Quantum energies with worldline numerics}

\author{H. Gies and K. Klingm\"uller}

\address{Institute for Theoretical Physics, Philosophenweg 16, D-69120
  Heidelberg} 

\ead{h.gies@, k.klingmueller@thphys.uni-heidelberg.de} 
\begin{abstract}
We present new results for Casimir forces between rigid bodies which
impose Dirichlet boundary conditions on a fluctuating scalar field. As
a universal computational tool, we employ worldline numerics which
builds on a combination of the string-inspired worldline approach with
Monte-Carlo techniques. Worldline numerics is not only particularly
powerful for inhomogeneous background configurations such as involved
Casimir geometries, it also provides for an intuitive picture of
quantum-fluctuation-induced phenomena. Results for the
Casimir geometries of a sphere above a plate and a new
perpendicular-plates configuration are presented.
\end{abstract}


\section{Introduction}

Casimir energies and forces are geometry dependent. Determining the
geometry dependence is a challenge both experimentally as well as
theoretically. Computing the geometry dependence of Casimir energies
can be viewed as a special case of the more general problem of
evaluating the effects of quantum fluctuations in a background
field $V(x)$. For instance, the space(-time) dependence of the
background field can then be used to model the Casimir geometry. 

A universal tool to deal with quantum fluctuations in background fields
is given by the effective action $\Gamma[V]$ which is the generating
functional for 1PI correlation functions for $V$. In the present work,
we consider a fluctuating real scalar quantum field $\phi$ interacting
with the background potential according to $\sim V(x) \phi^2$. For
this system, the effective action can be evaluated from 
\begin{eqnarray}
\Gamma[V]&=&
 -\ln \left( \int {\cal D}\phi\, \rme^{-\frac{1}{2} \int \phi
   (-\partial^2 +  m^2 + V)\phi} \right)\nonumber\\
&=& \quad \frac{1}{2}\sum_{ \lambda} \,\, 
\ln (\lambda^2 +m^2).
\label{eq:Gam1}
\end{eqnarray}
Here $\hbar$ and $c$ are set to 1.  The integral over the Gau\ss ian
fluctuations boils down to a sum over the spectrum $\{\lambda\}$ of
quantum fluctuations. This spectrum consists of the eigenvalues of the
fluctuation operator,
\begin{equation}
(-\partial^2+V(x))\,\phi =  \lambda^2\, \phi.\label{eq:SoQF}
\end{equation}
The relation to Casimir energies becomes most obvious by confining
ourselves to time-independent potentials $V(\mathbf{x})$. Then, the
sum over the time-like component of the spectrum can be performed. In
Euclidean spacetime, we use $-\partial^2=-\partial_t^2 -\nabla^2$,
$-\partial_t^2\to p_t^2$, and the summation/integration over $p_t$
results in
\begin{equation}
E[V]\equiv \frac{\Gamma[V]}{L_t}=\frac{1}{2} \sum \omega,\label{eq:EoV}
\end{equation}
where $\omega^2=\lambda^2-p_t^2$ denote the spatial
($p_t$-independent) part of the fluctuation spectrum. Here, we have
defined the Casimir energy from the effective action by dividing out
the extent $L_t$ of the Euclidean time direction. The relation to a
sum over ``ground-state energies'' $\sim \case{1}{2} \hbar \omega$ now
becomes obvious.

The general strategy for computing $E[V]$ seems straightforward:
determine the spectrum of quantum fluctuations and sum over the
spectrum. However, this recipe is plagued by a number of profound
problems: first, an analytic determination of the spectrum is possible
only in very rare, mainly separable, cases. Second, a numerical
determination of the spectrum is generally hopeless, since the spectrum
can consist of discrete as well as continuous parts and is generically
not bounded. Third, the sum over the spectrum is generally divergent,
and thus regularization is required; particularly in numerical
approaches, regularization can lead to severe stability problems. And
finally, an unambiguous renormalization has to be performed, such that
the physical parameters are uniquely fixed. 

For a solution of these problems, the technique of {\em worldline
numerics} has been developed \cite{Gies:2001zp} and has first been
applied to Casimir systems in \cite{Gies:2003cv}. This technique is
based on the string-inspired approach to quantum field theory
\cite{Schubert:2001he}. In this formulation, the (quantum mechanical)
problem of finding and summing the spectrum of an operator is mapped
onto a Feynman path integral over closed worldlines. For the present
scalar case, the effective action then reads
\begin{equation}
\Gamma[V]=
-\frac{1}{2}\int_{1/\Lambda^2}^\infty\frac{\rmd T}{T}\, \rme^{-m^2
  T}\, {\cal N} 
\int_{x(T)=x(0)} {\cal D} x\, 
\rme^{-\int_0^{T} \rmd\tau  
  \left( \frac{\dot{x}^2}{4} +
    V(x(\tau))\right)}.\label{eq:GamW}
\end{equation}
Now, the effective action, or, more specifically, the Casimir energy,
is obtained from an integral over an ensemble of closed worldlines in
the given background $V(x)$. This seemingly formal representation can
be interpreted in an intuitive manner: A worldline can be viewed as
the spacetime trajectory of a quantum fluctuation. The auxiliary
integration parameter $T$ is called the propertime and specifies a
fictitious ``time'' which the fluctuating particle has at its disposal
for traveling along the full trajectory. Larger values of the
propertime thus correspond to worldlines with a larger extent in
spacetime; hence, the propertime also corresponds to a smooth
regulator scale, with, e.g., short propertimes being related to
small-distance UV fluctuations.\footnote{For reasons of definiteness,
  we have therefore cut off the propertime integral at the lower bound
  at $1/\Lambda^2$ with the UV cutoff scale $\Lambda$.}

Most importantly from a technical viewpoint, the problem of finding
and summing over the spectrum is replaced by one single step, namely
taking the path integral. Moreover, for any given value of propertime
$T$ this path integral is finite. Possible UV divergencies can be
analyzed with purely analytical means by studying the small-$T$
behavior of the propertime integral, and thus no numerical
instabilities are introduced by the regularization procedure. The
renormalization can be performed in the standard way, for instance, by
analyzing the corresponding Feynman diagrams with the same propertime
regulator and by fixing the counterterms with the aid of
renormalization conditions accordingly. 

Further advanced methods which can deal with involved Casimir
configurations have been developed during the past years, each with
its own respective merits. In particular, we would like to mention the
semiclassical approximation \cite{semicl}, a functional-integral
approach using boundary auxiliary fields \cite{Golestanian:1998bx},
and the optical approximation \cite{Scardicchio:2004fy}. These methods
are especially useful for analyzing particular geometries by purely or
partly analytical means. 
 
Of course, a purely analytical evaluation of the path integral again
is only possible for very rare cases, but a numerical evaluation is
straightforwardly possible with Monte Carlo techniques and can be
realized with conceptually simple algorithms, as described in the next
section. Applications to Casimir geometries will be presented in
Sect.~\ref{sec:geom} and conclusions are given in Sec.~\ref{sec:conc}.

\section{Worldline numerics for Casimir systems}

With the aid of the normalization of the path integral
\cite{Gies:2001zp}, we note that \eqref{eq:GamW} can be written as
\begin{equation}
\Gamma[V]=-\frac{1}{2} \frac{1}{(4\pi)^2} \int_{1/\Lambda^2}^\infty
\frac{\rmd T}{T^3}\, \rme^{-m^2 T}\,\left( \left\langle \rme^{-\int_0^T \rmd
    \tau V(x(\tau))} \right\rangle_x -1\right), \label{eq:VEV}
\end{equation}
where the subtraction of $-1$ ensures that $\Gamma[V=0]=0$. The
expectation value in \eqref{eq:VEV} has to be taken with respect to
the worldline ensemble,
\begin{equation}
\langle \dots \rangle := \left(\int_{x(T)=x(0)} \mathcal D x \, \dots
\rme^{-\frac{1}{4} \int_0^T \rmd \tau {\dot x}^2} \right)
\left(\int_{x(T)=x(0)} \mathcal D x \
\rme^{-\frac{1}{4} \int_0^T \rmd \tau {\dot x}^2} \right)^{-1}.
\label{eq:VEV2}
\end{equation}
In the present work, we focus on the ``ideal'' Casimir effect induced
by real scalar field fluctuations obeying Dirichlet boundary
conditions; i.e., the boundary conditions are satisfied at infinitely
thin surfaces. This situation can be modeled by choosing $V(x)= g
\int_\Sigma \rmd\sigma\, \delta^{(4)}(x-x_\sigma)$, where $\rmd\sigma$
denotes the integration measure over the surface $\Sigma$, with
$x_\sigma$ being a vector pointing onto the surface. The Dirichlet
boundary condition is then strictly imposed by sending the coupling
$g$ to infinity, $g\to\infty$ \cite{Bordag:1992cm,Graham:2002xq}.

Moreover, we are finally aiming at Casimir forces between disconnected
rigid surfaces, which can be derived from the Casimir interaction
energy,
\begin{equation}
E_{\text{Casimir}}=E[V_1+V_2]-E[V_1]-E[V_2], \label{eq:ECas}
\end{equation}
where we subtract the Casimir energies of the single surfaces $V_1$
and $V_2$ from that of the combined configuration
$V_1+V_2$;\footnote{Alternatively, the single-surface subtraction
terms in \eqref{eq:ECas} can be viewed as subtracting the Casimir
energy for the surfaces at infinite separation.} the former do not
contribute to the force. This definition, together with the Dirichlet
boundary condition, leads to
\begin{equation}
E_{\text{Casimir}}=-\frac{1}{2} \frac{1}{(4\pi)^2} \int_{0}^\infty
\frac{\rmd T}{T^3}\, \rme^{-m^2 T}\,\left\langle\Theta_V[x]
\right\rangle_x , \label{eq:ECasW} 
\end{equation}
where $\Theta_V[x]=1$ if a given worldline intersects both surfaces
$\Sigma=\Sigma_1+\Sigma_2$ represented by the background potentials
$V=V_1+V_2$, and $\Theta_V[x]=0$ otherwise. This recipe has a simple
interpretation: any worldline which intersects both surfaces
corresponds to a quantum fluctuation that violates the Dirichlet boundary
conditions. Its ``removal'' from the set of all fluctuations
contributes ``one unit'' to the negative Casimir interaction energy.

The worldline numerical algorithm is based on an approximation of the
path integral by a sum over a finite ensemble of $n_{\text{L}}$ number
of paths, each of which is characterized by $N$ discrete points per
loop (ppl). These points are obtained by a discretization of the
propertime parameter on each loop, $x_i=x(\tau_i)$, $i=1, \dots, N$,
with $(x_i)_\mu\in\mathbbm{R}$. For an efficient generation of the
worldline ensemble which obeys a Gau\ss ian velocity distribution
required for \eqref{eq:VEV2}, various algorithms are available, see
\cite{Gies:2003cv,Gies:2005sb}.

In summary, worldline numerics offers a number of advantages: first,
the whole algorithm is independent of the background; no particular
symmetry is required. Second, the numerical cost scales only linearly
with the parameters $n_{\text{L}}$, $N$, and the dimensionality of the
problem.  The numerically most expensive part of the calculation is a
diagnostic routine that detects whether a given worldline intersects
both surfaces or not, returning the value $\Theta_V[x]=1$ or $0$,
respectively. Optimizing this diagnostic routine for a given geometry
can lead to a significant reduction of numerical costs. Details of
this optimization for the geometries considered below will be given
elsewhere \cite{HGKK}.

\section{Application to Casimir geometries}
\label{sec:geom}

\subsection{Sphere above plate}

The geometry of a sphere above a plate is the most relevant
configuration as far as recent and current experiments are concerned
\cite{Lamoreaux:1996wh}. Therefore, also worldline numerics has first
been applied to this case \cite{Gies:2003cv}. Here we extend these
studies, arriving at significantly improved results with much smaller
error bars and for a wider range of parameters. In the following, we
exclusively discuss the massless case, $m=0$. 

It is interesting to compare our results to the proximity force
approximation (PFA) \cite{pft1} which is the standard tool for estimating
the effects of departure from planar geometry for Casimir effects. In
this approach, the curved surfaces are viewed as a superposition of
infinitesimal parallel plates, and the interaction energy is then
obtained by
\begin{equation}
E_{\text{PFA}}=\int_{\Sigma_{\text{PFA}}} E_{\text{PP}}(d)\,
\rmd\sigma. \label{eq:EPFA}
\end{equation}
Here, $\Sigma_{\text{PFA}}$ denotes a ``suitable'' auxiliary surface
in between the Casimir surfaces $\Sigma_1$ and $\Sigma_2$, and
$d\sigma$ is the corresponding surface element of
$\Sigma_{\text{PFA}}$. The distance $d$ between two points on
$\Sigma_1$ and $\Sigma_2$ has to be measured along the normal to
$\Sigma_{\text{PFA}}$. Obviously, the definition of $E_{\text{PFA}}$
is ambiguous, owing to possible different choices of
$\Sigma_{\text{PFA}}$. The two extreme cases are
$\Sigma_{\text{PFA}}=\Sigma_1$ or $\Sigma_{\text{PFA}}=\Sigma_2$. The
difference in $E_{\text{PFA}}$ for these two cases is considered to
represent a rough error estimate of the PFA.  In the above formula,
$E_{\text{PP}}$ denotes the classic parallel-plate result for the
energy per unit area $A$ \cite{Casimir:dh},
\begin{equation}
\frac{E_{\text{PP}}(a)}{A}=-c_{\text{PP}}\frac{\pi^2}{1440}
\frac{1}{a^3}, 
\label{eq:EPP}
\end{equation}
with $a$ denoting the plate separations, and $c_{\text{PP}}=2$ for an
electromagnetic (EM) field or a complex scalar, and $c_{\text{PP}}=1$
for the present case of a real scalar field fluctuation.

\begin{figure}[t]
\begin{center}
{\unitlength=1mm
\begin{picture}(80,80)
\put(0,0){
\epsfig{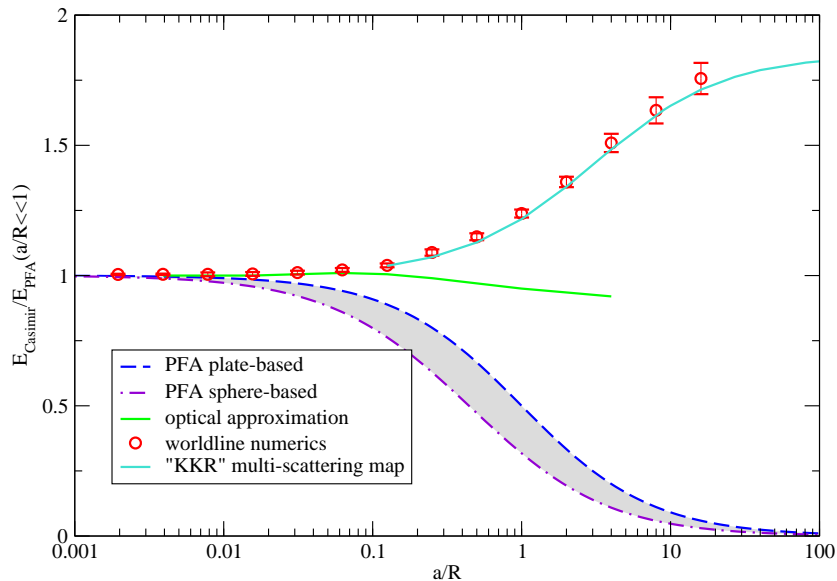} 
}
\end{picture}
}
\end{center}

\vspace{-0.3cm}
\caption{Casimir energy for the sphere-plate configuration normalized
  to the zeroth-order PFA formula \eqref{eq:Enorm}: the dashed and
  dot-dashed lines depict the plate-based and the sphere-based PFA
  estimates, respectively. The circle symbols display our worldline
  numerical result. The deviation from the PFA estimate characterize
  the relevance of Casimir curvature effects. Also shown is the result
  from the optical approximation \cite{Scardicchio:2004fy}, which,
  within its validity limits $a/R\lesssim0.1$, agrees well with our
  result. For larger $a/R$, we find satisfactory agreement with the "KKR"
   multi-scattering map method presented at this workshop \cite{Wirzba}.}
\label{fig:sphereplate}
\end{figure}

For the configuration of a sphere above a plate, we can choose
$\Sigma_{\text{PFA}}$ equal to the plate (plate-based PFA), or equal
to the sphere (sphere-based PFA) as the extreme cases. The
corresponding results of the integral of \eqref{eq:EPFA} can be given
in closed form, see, e.g., \cite{Scardicchio:2004fy}.  In the limit of
small distances $a$ compared to the sphere radius $R$, $a/R\ll 1$,
both PFA's agree,
\begin{equation}
E_{\text{PFA}}(a/R\ll1)= -c_{\text{PP}}\frac{\pi^3}{1440}
\frac{R}{a^2}. \label{eq:Enorm}
\end{equation}
It is useful to display the resulting Casimir energies normalized to
this zeroth-order small-separation PFA limit as it is done in figure
\ref{fig:sphereplate}. The dashed and dot-dashed lines depict the
plate-based and sphere-based PFA, respectively. For larger
separations, both cases predict a decrease of the Casimir energy
relative to the zeroth-order PFA in \eqref{eq:Enorm}.

Our numerical worldline estimate confirms the zeroth-order PFA in the
limit of small $a/R$. But in contrast to the PFA curves, worldline
numerics predicts a relative increase of the Casimir energy in
comparison with \eqref{eq:Enorm} for larger separations/smaller
spheres. We conclude that the PFA should not at all be trusted beyond
the zeroth order: the first-order correction does not even have the
correct sign. As a most conservative estimate, we observe that the PFA
deviates from our result by at least 1\% for $a/R>0.01$, which
confirms and strengthens the result of \cite{Gies:2003cv}. A more
detailed investigation of the validity bounds of the PFA will be given
elsewhere \cite{HGKK}. It is interesting to observe that the
zeroth-order PFA \eqref{eq:Enorm} still seems to be a reasonable
estimate up to $a/R\simeq 0.1$, indicating that the true curvature
effects compensate for the higher-order PFA corrections.

Finally, we note that our results agree with the optical approximation
\cite{Scardicchio:2004fy} for $a/R\lesssim 0.1$, confirming the
absence of diffractive effects in this regime, which are neglected by
the optical approximation. For even larger separations $a$, we observe
a monotonous increase of the Casimir energy relative to
\eqref{eq:Enorm}. In this regime, our results agree quantitatively
with those obtained from the ``KKR'' multi-scattering map method
presented by A. Wirzba at this QFEXT05 workshop \cite{Wirzba}. Most
importantly, we do not observe a Casimir-Polder law for large $a/R$,
which would manifest itself in an $(a/R)^{-2}$ decrease in figure
\ref{fig:sphereplate} at the large-$(a/R)$ side. Since a
Casimir-Polder law is expected for the electromagnetic case, our
results for the Dirichlet scalar provide clear evidence for the fact
that the relation between Casimir forces for the EM field and for the
Dirichlet scalar is strongly geometry dependent. In fact, the
latest results of Ref.[14] include an analytic proof that the energy
ratio of Fig. \ref{fig:sphereplate} approaches a constant,
$180/\pi^4\simeq 1.848$ for large $a/R$ for the Dirichlet scalar.  The
analysis of the sphere-plate configuration for the EM field therefore
still remains an open unsolved problem. Note that this does not affect
our conclusions about the PFA, since also the PFA does not treat the
EM case or the Dirichlet scalar in a different manner.

\subsection{Perpendicular plates}

\begin{figure}[t]
{\unitlength=1mm
\begin{picture}(160,75)
\put(0,0){
\epsfig{figure=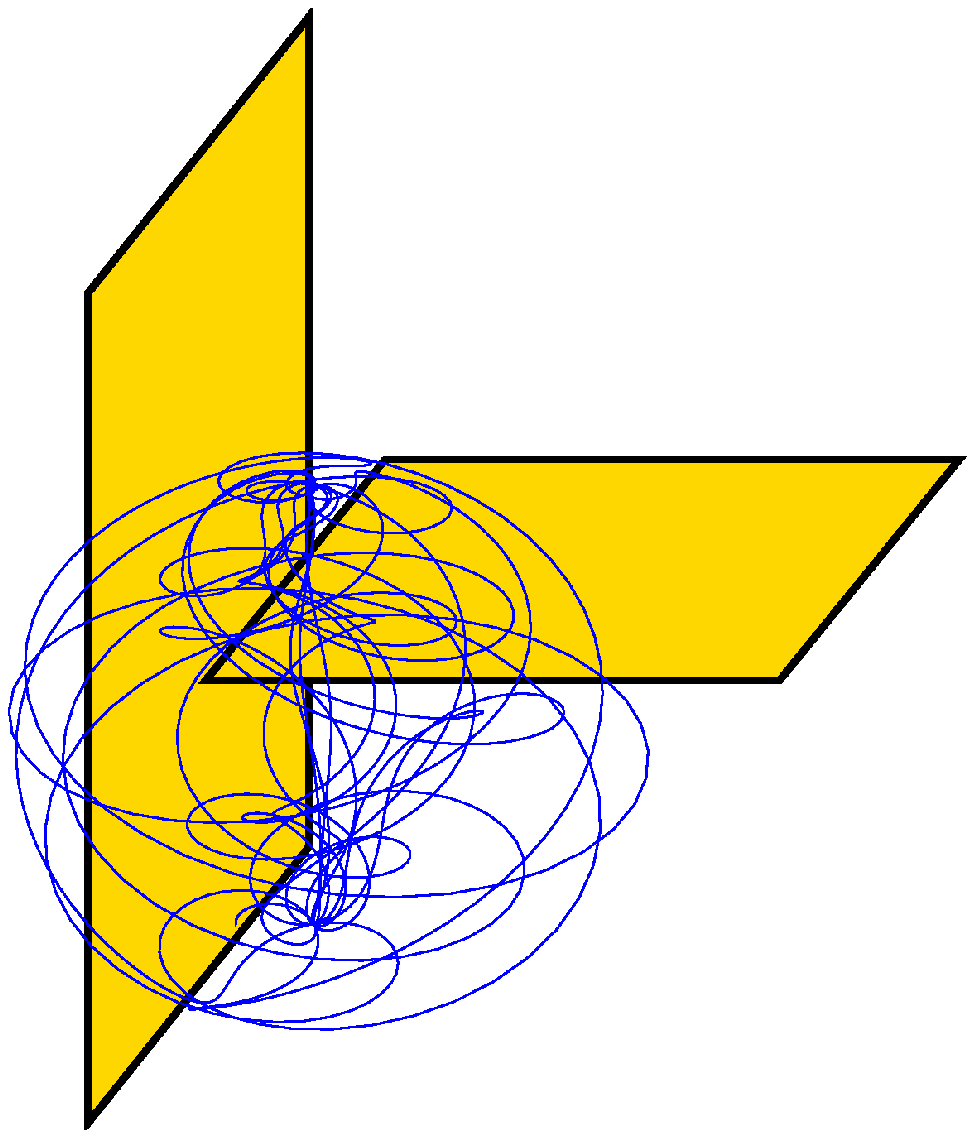,width=7cm}
}
\put(70,0){
\epsfig{figure=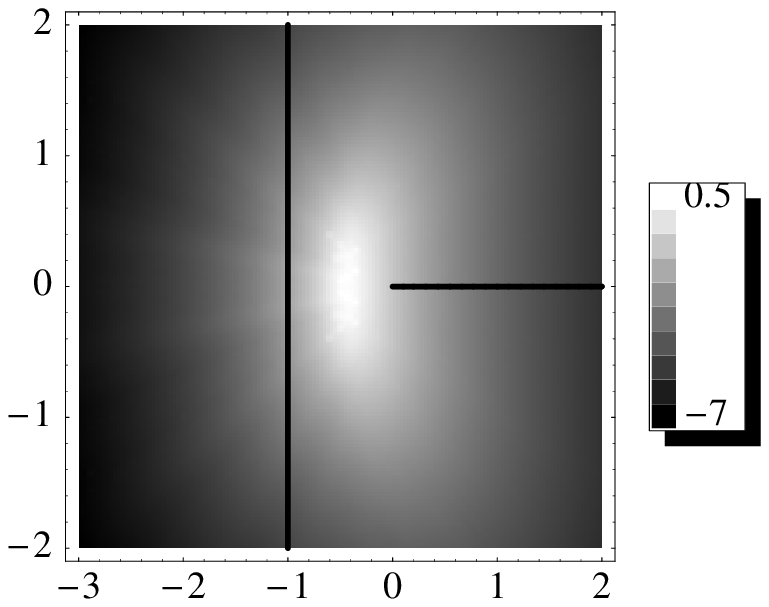,width=9cm}
}
\end{picture}
}
\vspace{-0.3cm}
\caption{Left panel: sketch of the perpendicular-plates configuration
  with (an artist's view of) a typical worldline that intersects both
  plates. Right panel: density plot of the effective action density
  $\mathcal L$ for the perpendicular plates case; the plot shows $\ln
  (2 (4\pi a^2)^2 |\mathcal L|)$. The position of the perpendicular
  plates are indicated by solid lines for illustration.}
\label{fig:density}
\end{figure}

A particularly inspiring geometry is given by a variant of the classic
parallel-plate case: a semi-infinite plate perpendicular to an
infinite plate, such that the edge of the semi-infinite plate has a
minimal distance $a$ to the infinite plate, see figure
\ref{fig:density} (left panel). Whereas Casimir's parallel-plate case
has only one nontrivial direction (the one normal to the plates), this
perpendicular-plates case has two nontrivial directions but still only
one dimensionful scale $a$. This fixes the scale dependence of the
energy per unit length unambiguously,
\begin{equation}
\frac{E_{\bot}(a)}{L_{\text{T}}}=-\gamma_{\bot}\frac{\pi^2}{1440}
\frac{1}{a^2}, 
\label{eq:Ebot}
\end{equation}
where $L_{\text{T}}$ denotes the extent of the system along the remaining
trivial transversal direction. The unknown coefficient $\gamma_\bot$
results from the effect of quantum fluctuations in this geometry and
will be determined by worldline numerics.

Let us first note that the PFA does not appear to be useful for the
perpendicular-plates case, because the surfaces cannot reasonably be
subdivided into infinitesimal surface elements facing each other from
plate to plate. For instance, choosing either of the plates as the
integration surface in \eqref{eq:EPFA}, the PFA would give a zero
result. However, in the worldline picture, it is immediately clear
that the interaction energy is nonzero, because there are many
worldlines which intersect both plates, as sketched in figure
\ref{fig:density} (left panel). As a direct evidence, we plot the
negative effective action density $\mathcal L$ (effective Lagrangian)
in figure \ref{fig:density} (right panel); the effective action is
obtained from $\Gamma= \int \rmd^4 x \mathcal L$. Brighter areas
denote a higher density of the center-of-masses of those worldlines
which intersect both plates.

Integrating over the effective action density, we obtain the universal
coefficient
\begin{equation}
\gamma_{\bot}=0.87511 \pm 0.00326,
\end{equation}
using $n_L=40000$ worldlines with $N=200000$ ppl generated by the {\em
  v loop} algorithm \cite{Gies:2003cv}.

\section{Conclusions}
\label{sec:conc}

We have presented new results for interaction Casimir energies, giving
rise to Casimir forces between rigid bodies, induced by a fluctuating
real scalar field that obeys Dirichlet boundary conditions. We have
used worldline numerics as a universal tool for dealing with quantum
fluctuations in inhomogeneous backgrounds. 

For the experimentally relevant sphere-plate configuration, we have
performed extensive numerical studies, confirming earlier findings
\cite{Gies:2003cv} with a significantly higher precision and narrowing
the validity bounds of the proximity force approximation even
further. Moreover, our results for small spheres for the Dirichlet
scalar shows no sign of a Casimir-Polder law, as it would be expected
for the EM field. This provides clear evidence for a different role of
Casimir curvature effects for these two different field theories,
leaving the sphere-plate configuration with a fluctuating EM field as
a pressing open problem.

Furthermore, we have investigated a new geometry of two perpendicular
plates which has been inaccessible so far for other approximation
techniques. The configuration is representative for a whole new class
of Casimir systems involving sharp edges, where diffractive portions
of the fluctuating field will play a major role.

\ack
It is a pleasure to thank Emilio Elizalde and his team for the
organization of this workshop and for creating such a stimulating
atmosphere. H.G. acknowledges useful discussions with G.V. Dunne,
T. Emig, A. Scardicchio, O. Schr\"oder, A. Wirzba, and H. Weigel. This
work was supported by the Deutsche Forschungsgemeinschaft (DFG) under
contract Gi 328/1-3 (Emmy-Noether program) and Gi 328/3-2.

\end{document}